# Java Card For PayTV Application

Pallab Dutta
Team Leader, Centre For Development Of Telematics
Electronic City, Bangalore, India.
Email: pallab@cdot.in

*Abstract*— Smart cards are widely used along with PayTV receivers to store secret user keys and to perform security functions to prevent any unauthorized viewing of PayTV channels. Java Card technology enables programs written in the Java programming language to run on smart cards. Smart cards represent one of the smallest computing platforms in use today. The memory configuration of a smart card are of the order of 4K of RAM, 72K of EEPROM, and 24K of ROM. Using Java card provides advantages to the industry in terms of ease of coding, faster time to market and faster upgrades as compared to plain smart cards . Also different applications like payTV, e-commerce, health-card can easily be implemented in a single java card as multiple applets corresponding to each application can coexists in a single java card. But there are security concerns in java cards and also the performance issues. In this paper, we analyse the suitability of using Java card for PayTV applications as part of conditional access system in place of plain smart cards.

I. INTRODUCTION

Satellite PayTV network consists of a head-end, uplinking transmission system, satellite and reception system along with subscriber Management system (SMS) and subscriber Authorization System (SAS). At the head-end (transmit side), the digital content (including video, audio and data), which the operator wishes to restrict access, is scrambled (DVB-CSA) by the control word (CW) derived from a constantly changing pseudo-random binary sequence generator. The control word also needs to be protected: the control word is encrypted with a service key ( also called authorization key ) (SK / AK) . The encrypted control word is then packaged into entitlement control message (ECM). Then, the service key is encrypted with the individual key (IK) supplied by the subscriber management system (SMS) and is then packaged with entitlement data into entitlement management message (EMM). Finally, the scrambled content, entitlement control message, and entitlement management message are together broadcasted in the same channel. This is shown as in Figure 1.

At the reception-end, the set-top box (STB) and smart card authenticate each other. STB filters entitlement management message and entitlement control message according to the parameters provided by the smart card (SC) and then forwards these messages to smart card. Smart card decrypts entitlement management message using individual key stored in smart card to get service key/authorization key and the entitlement data. After having passed the verification of the access entitlement, smart card uses the service key to decrypt the encrypted control word and returns the control word towards set-top box so that set-top box will be allowed to descramble the scrambled content. This is shown in the Figure 2. The smart card and part of STB where entitlement messages are filtered and smart card handshake takes place is called conditional access sub system.

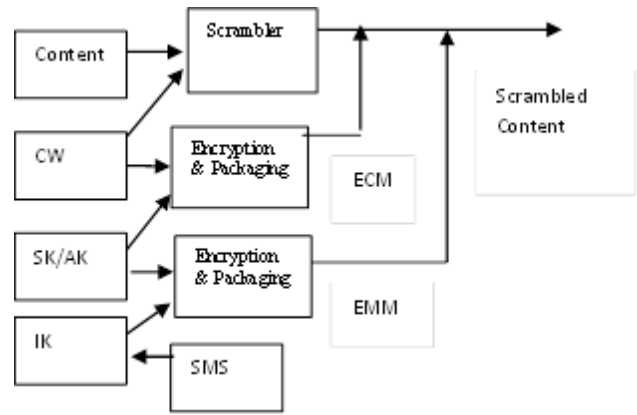

Figure 1. PayTV Headend

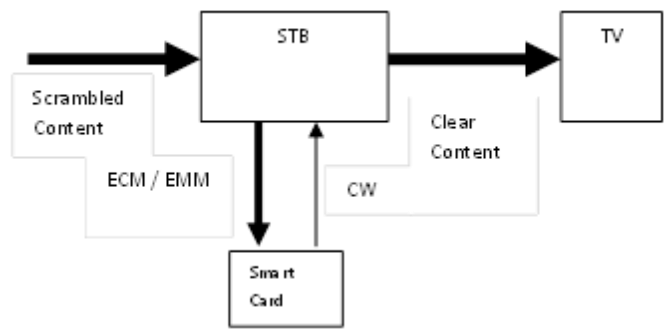

Figure 2. PayTV Receiver

Java Card technology enables programs written in the Java programming language to run on smart cards. A JavaCard is a typical smart card: it conforms to all smart card standards and thus requires no change to existing smart card-aware applications. However, JavaCard has a specific important feature that makes it unique: a Java Virtual Machine is implemented in its read-only memory (ROM) mask. The JVM controls the access to all smart card resources, such as memory and I/O, and thus essentially serves as the smart card's operating system. The JVM executes a Java bytecode subset on the smart card, ultimately providing the functions accessible from outside, such as signature, log-in, and loyalty applications. Java Card offers definite advantages over plain Smart cards in terms of ease of programming, faster time to market, coexistence of multiple applications, interoperability etc. In this paper we analyse the suitability of using java card for PayTV applications as part of conditional access system in place of plain smart cards. In Section II, we list the advantages and disadvantages of Java Card over conventional Smart card. Section III analyses java card technology in detail from architecture and security point of view. In Section IV, we give the requirements to be met by java card for using in PayTV application. In Section V, we have given the experimental setup and performance details of java card. Section VI, gives the conclusion.

## II. ADVANTAGES AND DISADVANTAGES OF JAVA CARD

*A. Advantages of Java Card*

There are several advantages of java card over plain smart cards. The following are the main advantages of Java Cards:

Interoperable: Applets developed with Java Card technology runs on any Java Card technology-based smart card, independently of the card vendor and underlying hardware.

Secure: Java Card technology relies on the inherent security of the Java programming language to provide a secure execution environment. It was designed through an open process, and the platform's proven industry deployments and security evaluations ensure that card issuers benefit from the most capable and secure state of the art technology.

Multi-Application Capable**:** Java Card technology enables multiple applications to co-exist securely on a single smart card. Hence same card can be used for several end applications.

Dynamic: New applications can be installed securely after a card has been issued, providing card issuers with the ability to dynamically respond to their customer's changing needs.

Open**:** Java Card application developers benefit from object-oriented programming and design, and have access to off-the-shelf Java development tools.

Faster Development: Because of Java based development environment, this gives the advantage of faster time to market and also debugging easier and more error free code can be developed faster compared to a native application development.

Compatible with Existing Standards: The Java Card API is compatible with international standards for smart cards such as ISO7816, or EMV. It is referenced by major industry-specific standards such as Global Platform and ETSI.

*B. Disadvantages of Java Card*

Although java card offers lot of advantages, it also has a few disadvantages; following are the main disadvantages of Java card:

Less Memory Space: Java Card Run time Environment occupies extra memory space in the java card, hence the memory space available for the user application reduces.

Slower: Java code execution is normally slower than the native code written in middle level or low level programming languages.

Security: Because java card is an open system, security through obscurity does not workout.

## III. JAVA CARD TECHNOLOGY AND SECURITY

*A. Java Card Technology*

Before going into security issues let us first discuss about Java Card Architecture.

As shown in the figure-3, the low level operations like physical data transfer, file management, communication and instruction execution are handled by microcontroller program and operating system.

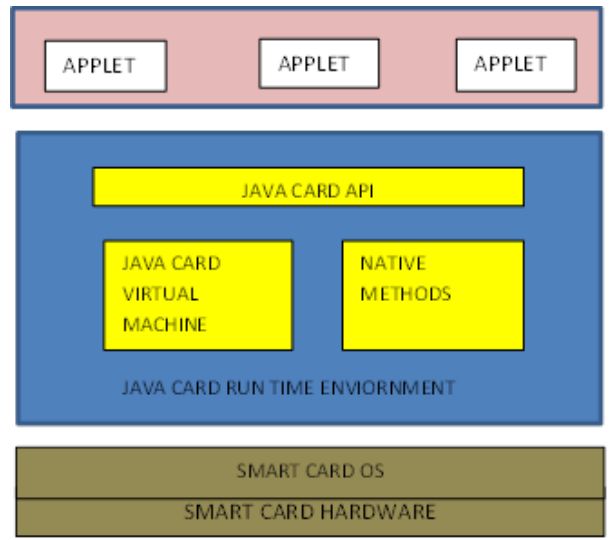

Figure 3. JAVA Card Architecture

JCRE (Java Card Runtime Environment) sits over this layer and it consists of Java Card API, JCVM (Java Card Virtual Machine) and native methods. The native methods are platform dependent and they directly interact with underlying operating system, it provides services like cryptography, memory allocation and I/O operations. Java Card API also called as framework consists of all Java packages required for various Java Card functionalities. JCVM executes the Java Byte Code just like JVM but the difference is that JCVM is split into two parts. One is Java Card Converter which is off-card, converts *.class* files into card compatible format (*.cap* files) and the other is on-card part which interprets the byte code. JCRE and operating system program are stored into ROM at the time of manufacture of chip while card applications are loaded into EEPROM.

Java Card Language Subset: Because of its small memory footprint, the Java Card platform supports only a carefully chosen, customized subset of the features of the Java language. This subset includes features that are well suited for writing programs for smart cards and other small devices while preserving the object-oriented capabilities of the Java programming language. Table 1 highlights some notable supported and unsupported Java language features. Keywords of the unsupported features are also omitted from the language. Many advanced Java smart cards provide a garbage collection mechanism to enable object deletion.

TABLE 1: Supported and Unsupported Java Card features

| Supported Java feature | Unsupported Java Feature |
|---|---|
| - Small primitive data types: boolean, byte, short<br>- One-dimensional arrays<br>- Java packages, classes, interfaces, and exceptions<br>- Java object-oriented features: inheritance, virtual methods, overloading and dynamic object creation, access scope, and binding rules<br>- The int keyword and 32-bit integer data type support are optional. | - Large primitive data types: long, double, float<br>- Characters and strings<br>- Multidimensional arrays<br>- Dynamic class loading<br>- Security manager<br>- Garbage collection and finalization<br>- Threads<br>- Object serialization<br>- Object cloning |

Java Card Application Deployment: Java card application is first compiled using Java compiler and class files are obtained. These class files are converted into cap files (Converted Applet files) and then installed on to the card as shown in the figure 4. Now applet must be installed and registered in the card. This is done either by the installation program sitting on JCRE or by the *install* method of abstract applet class. When the installation and registration is successful, an applet is instanced and all objects related to it are created.

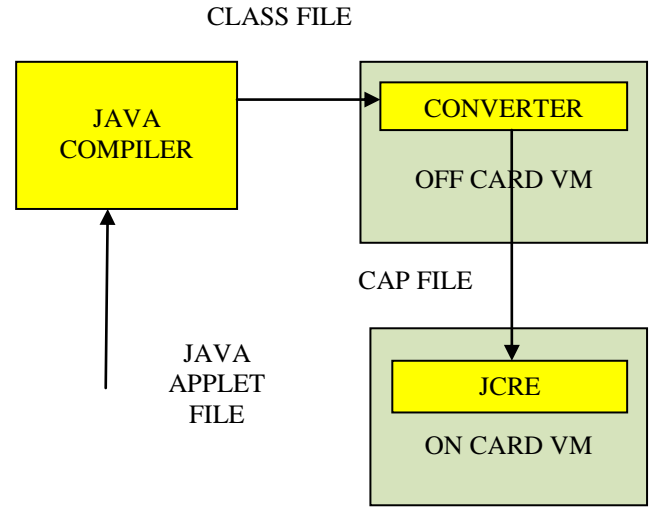

Figure 4. Java Card Application Deployment

### B. Java Card Security

#### 1. Java card security features:

Java Card provides lot of security features like applet firewalls, Java type safe, object sharing etc. As smart card uses cryptography technique for data encryption, Java Card provides Crypto API. We discuss about the features in detail below:

Prohibition of Dynamic Class Loading: Java Card does not allow dynamic class loading. This means that Java classes are not be downloaded on-the-fly during execution; a security risk assumed by standard Java. Instead, Java applets are usually burned in during masking when the card is initialized and personalized prior to being fielded.

Applet Firewall: Java Card can handle multiple applets and these applets can be from different venders so it is not secured if objects of one applet are accessible to another applet. Java Card uses applet firewall and type safe feature of Java to restrict unauthorized access of data. That is Java Card defines context for each applet which represents all the objects accessible to a particular applet.

Object Sharing: Java Card also provides a facility to share objects by implementing *Sharable* interface. This allows the applets to share objects and at the same time it restricts them from accessing methods which are accessible only to the owner of the object that is remote method invocation is not allowed.

Java Card API: The Java Card APIs consist of a set of customized classes for programming smart card applications according to the ISO 7816 model. The APIs contain three core packages and one extension package. The three core packages are java.lang, javacard.framework, and javacard.security. The extension package is javacardx.crypto. Many Java platform classes are not supported in the Java Card APIs. For example, the Java platform classes for GUI interfaces, network I/O, and desktop file system I/O are not supported. The reason is that smart cards do not have a display, and they use a different network protocol and file system structure. Also, many Java platform utility classes are not supported, to meet the strict memory requirements. The classes in the Java Card APIs are compact. They include classes adapted from the Java platform for providing Java language support and cryptographic services. They also contain classes created especially for supporting the smart card ISO 7816 standard.

*2. Security Risks in Java Card*

Although java card offers lot of security features, it also has few security risks. The following are the main risks associated with java cards.

Absence of Garbage Collection: In Java, Garbage Collection feature controls the freed and unused objects of the applet but this feature is absent in Java Card (because of its memory and processor resource limitations) which may cause serious threat like handle of a free object is allocated to a new object. If any error occurs in the program then the memory allocated to the object is never freed which lead to memory leakage. This is a serious problem especially on smart card with such limited memory resources and application may get hanged up in the middle of transaction and suffer from denial-of-service attack.

Native methods: JCRE interacts with card resources using native methods which are mostly written in C or C++, card vender can upgrade card by adding new native methods and applets can use these methods. Java card firewalls have no control over these native methods so all the firewall security fails when it comes to native methods.

Exception Handling: Another security threat is with exception handling, it can be controlled if applets are intensively tested using testing tools before loading on card. If an exception is not handled in any of the called methods it becomes unpredictable and may hang the applet and corrupt the card.

Protocol Interactions: In today's smart card market, multi-application cards are becoming very common and also convenient. That is, rather then carrying several different cards for different purposes, the applications are integrated on a single card. For example, a smart card may provide driver's license or social security identification, a credit line, a debit account, e-cash, health insurance information, calling card applications, digital certificates for authentication, and different loyalty programs. Insofar as multiple applications implement different protocols (many of which may require authentication), unintended protocol interactions may compromise the security of the card and its multiple applications. Because of small memory constraints in smart cards, it is currently impractical to store unique cryptographic public/private key pairs for each different application requiring cryptographic functions. Because of this, multiple applications often share the same key material. Different protocols that share the same key material can introduce potential security problems that would not otherwise exist with each protocol considered in isolation. That is, the security of two different protocols do not necessarily compose, especially given protocol interactions.

Physical Security: The physical security risks of Java cards are same as those of plain smart cards. Various physical attacks like, DPA (differential Power Analysis) need to be countered measured in Java Cards in a similar way that of plain smart cards.

IV. REQUIREMENTS TO BE MET BY JAVA CARDS FOR USING IN PAYTV APPLICATION

*A. Security of Cryptographic Algorithms*

In PayTV applications, both symmetric and asymmetric algorithms are used. Although, PayTV standards does not mandate usage of any specific cryptographic algorithm for key distribution and scrambling, but generally two layer security scheme is employed as mentioned in the section I of this paper. Symmetric cryptographic algorithm like AES and Triple DES are used. An asymmetric algorithm like RSA is used. Pseudo Random numbers are required to be generated as part of authentication protocol used as session keys between smart card and STB. The implementation of complex crypto algorithm shall be easier and faster through standard APIs.

*B. Confirmation to ISO7816 standard*

The Smart Card – PayTV receiver interface is ISO 7816, hence the Java card shall also confirm to this specification. The Java Card APIs consist of a set of customized classes for programming smart card applications according to the ISO 7816 model. The classes in the Java Card APIs are compact. They include classes adapted from the Java platform for providing Java language support and cryptographic services. They also contain classes created especially for supporting the smart card ISO 7816 standard.

*C. High Security:*

In the payTV scenario, if the control word (CW) is cracked, then the TV channels can be decoded unauthorised. This will lead to huge revenue loss to the operator. The Smart Card and the receiver – any one of these two may be malicious; hence requires bi-directional authentication so that the receiver is able to verify the identity of the Smart Card and also Smart Card is able to verify the identity of the receiver. Otherwise, a cloned smart card will be able to decode the CW or a fake receiver will be able to store secret information from a genuine Smart Card and create large number of fake Smart Cards. It shall be possible to implement bi-directional authentication in Java Card in an efficient way both in terms of memory usage and time required for executing complex crypto algorithm.

*D. Performance*

The Java card to be used in PayTV receiver need to satisfy stringent performance requirements. In payTV system, Control word usually changes in every 10 seconds, also control word is per channel. For better user experience, ECM is transmitted every 100ms, so Smart card need to decode the ECM message and extract Control Word and send it to the receiver via ISO 7816 interface, once in every 10s minimum. Generally Control Word extraction and transmitting it to the receiver involves triple DES or AES encryption / decryption. When the receiver is switched on or Smart Card is inserted in the receiver, bi-directional authentication takes place between the smart card and the receiver. This involves RSA decryption, Random Number Generation and AES decryption. This phase has to be completed within 2-3 seconds for smooth user experience.

as part of conditional access system. As per Kerckhoff's principle, an well-known principle in cryptography, security though obscurity does not give overall system security in long run, hence an open smart card platform like Java Card platform, does not compromise the security of a proprietary smart card system rather it may enhance the security. Also as we have seen from the basic performance data, the triple DES, AES and RSA gives the acceptable performance for payTV applications. However in some practical payTV systems, there will be additional computational load wrt certificate verification for mutual authentication, dynamic session key generation etc., and in cases, it is required to implement Schnorr Algorithm. This will require modulo multiplication operation and not be possible to use the RSA co-processor directly. In such cases a proper mechanism to use the individual building blocks of the co-processor need to be designed and made known to the Java Card user to use the computational power of the co-processor to achieve acceptable performances in such cases also.

## V. EXPERIMENTAL SETUP AND PERFORMANCE

The experimental setup consists of a core2 duo PC acting as PayTV receiver, a smart card reader is connected to the USB port of the PC. JCOP 21 Java card is used to perform the cryptographic functions triple DES, AES, RSA. Eclipse IDE is used along with JCOP plugins for the Java Card Applet Development. The Java card used here has triple DES, AES and RSA coprocessors. The EEPROM size is 72K and RAM size is 4096 B. Applets are loaded in EEPROM. 128 bit key length for AES and 1024 bit key of RSA is used.

It is observed that Triple DES encryption is taking around 5 ms, AES is taking around around 10 ms. RSA takes around 100 ms. As we have seen in the performance requirements, the CW changes in every 10 second and ECM is repeated per 100ms, hence the Triple DES and AES performances are acceptable for PayTV application. Also RSA is not used runtime for CW extraction and RSA is used only when smart card is inserted in the receiver or when the receiver is turned on with the smart card, hence the RSA performance is also acceptable for the given application of payTV conditional access system.

## VI. CONCLUSIONS

We have discussed the payTV conditional access system requirements and analysed the advantages and disadvantages of Javacard in general and also in the specific context of payTV receiver. We have seen that although there are some disadvantages of Java Cards but those do not create any hindrance for the Java Cards to be used in the payTV receivers

[14] ISO/IEC 7816-4: Information technology – Identification cards – Integrated circuit(s) cards with contacts – Part 4: Inter industry commands for interchange, International Organization for Standardization, 1997.

AUTHORS PROFILE

**Mr.Pallab Dutta** is working in C-DOT (Centre For Development Of Telematics) for last 16 years. He has worked in various fields/projects of Telecommunication during this tenure. His area of interests are Embedded system design, High Performance Computing, Cryptography, Multi-Agent system etc. He Did BTech from REC (presently known as NIT) Calicut in ECE in 1997. He completed his ME (CSE) in 2007 and presently pursuing his PhD in CSE.